%% file: axion_prd_v2.tex
\documentclass[prd,twocolumn,nofootinbib,notitlepage]{revtex4-1}
\input{def.tex}

\usepackage{enumitem}
\usepackage{graphicx}
\usepackage{subfigure}
\usepackage{amssymb}
\usepackage{multirow}
\usepackage{amsmath}
\usepackage{bbm}
\usepackage{color}
\usepackage[retainorgcmds]{IEEEtrantools}
\begin{document}
\title{Constraining axionlike particles using the distance-duality relation}
\author{Prabhakar Tiwari}
\date{\today}
\email{ptiwari@physics.technion.ac.il}
\affiliation{ Physics Department and the Asher Space Research Institute - Technion, Haifa 32000, Israel\\}
\begin{abstract}
One of the fundamental results used in observational cosmology is the distance duality relation (DDR),
which relates the luminosity distance, $\dl$, with angular diameter distance, $\da$, at a given redshift $z$. 
We employ the observed limits of this relation to constrain the coupling of axion like particles (ALPs) 
with photons. With our detailed $3D$ ALP-photon mixing simulation in standard $\Lambda$CDM universe and latest 
DDR limits observed in  Holand \& Barros (2016) we limit the coupling constant 
$g_{\phi}\le 6\times10^{-13} \gev^{-1}\left(\frac{nG}{\langle B \rangle_{\rm Mpc}}\right)$ for ALPs of mass $\le 10^{-15}$ eV. 
The DDR observations can provide very stringent constraint on ALPs mixing in future.
Also any deviation in DDR can be conventionally explained as photons decaying to axions or vice-versa. 
\end{abstract}
\maketitle
\section{Introduction}
\label{intro}
The many extensions of Standard Model (SM) predict the existence of new  light particles 
which may weakly couple with visible matter and photon. The famous known example is axion or 
axion like particles (ALPs) which arise naturally as Goldstone bosons in spontaneously broken global symmetries 
\cite{Peccei:1977prl,Peccei:1977prd,McKay:1977,Weinberg:1978,Wilczek:1978,
McKay:1979,Kim:1979,Dine:1981,Kim:1987}. The ALPs are massless if the broken symmetry was exact, however, if 
it was approximate, the ALPs are light pseudoscalar. These particles, if exist, will potentially 
show many interesting cosmological as well as astrophysical signatures. Axions are also dark matter candidate.
Interestingly, the ultra-light axion mixed dark matter has been proposed to resolve the missing satellites problem, 
the cusp-core problem and the `too big to fail' problem \citep{Marsh:2014}. This solely gives strong 
motivation to study ultra light axions. The axion coupling with photons modify almost all the cosmological 
observations, and  many such effects have been widely studied in literature 
\cite{Harari:1992,Mohanty:1993,Das:2001,Kar:2002,Kar:2002cqg,Csaki:2002,
Csaki:2002prl,Grossman:2002,Jain:2002vx,Song:2006,Mirizzi:2005,Raffelt:2008,Gnedin:2007,Mirizzi:2007,
Mirizzi:2009,Finelli:2009,Agarwal:2012,Ostman:2005,Lai:2006,Hooper:2007,Hochmuth:2007,Chelouche:2009,
Agarwal:2012,Tiwari:2012ax}. 

The coupling of ALPs with photons in presence of external magnetic field 
\cite{Clarke:1982,Sikivie:1983,Sikivie:1985,Sikivie:1988,Maiani:1986,Raffelt:1988,Carlson:1994,Bradley:2003,
Das:2004qka,Das:2004ee,Ganguly:2006,Ganguly:2009} has interesting consequences, for example the ALP can convert into 
a photon and vice versa i.e. the number of photons is not conserved. Also, the mixing introduces 
polarization in radiation \cite{Harari:1992,Mohanty:1993,Das:2001,Kar:2002,Kar:2002cqg,
Csaki:2002,Csaki:2002prl,Grossman:2002,Jain:2002vx,Song:2006,Mirizzi:2005,Raffelt:2008,Gnedin:2007,
Mirizzi:2007,Finelli:2009,Agarwal:2012,Ostman:2005,Lai:2006,Hooper:2007,Hochmuth:2007,Chelouche:2009,
Agarwal:2012,Tiwari:2012ax}. Due to the weak coupling of axions with photons, the change in intensity and polarization 
is very small for light from local sources. However, it becomes significant for 
light coming from far cosmological sources. As a result the radiation from distant cosmological sources has been
employed to constrain axion-photon mixing by several authors e.g. from Cosmic Microwave Background Radiation (CMBR) 
we have strong limits on coupling for a wide range of axion mass \citep{Mirizzi:2005,Agarwal:2008ac,Tiwari:2012ax}. 
Furthermore, the axion-photon mixing has been applied to explain several physical puzzling observations, for example 
to explain large scale polarization alignment in distant quasars \cite{Hutsemekers:1998,Hutsemekers:2001fv, Hutsemekers:2005iz,Hutsemekers:2008, 
Agarwal:2011,Agarwal:2012}, polarization properties of radio galaxies \cite{Harari:1992,Jain:2002vx,Jain:1998r} and 
the dimming of distant supernovae \cite{Csaki:2002,Jain:2002vx,Mirizzi:2005}. In addition there are various experiments looking for direct and indirect 
observations of axions \cite{Mohanty:1993,Raffelt:2008,Dicus:1978,Dearborn:1986,Raffelt:1987,Raffelt:1988prl,
Turner:1988,Janka:1996,Keil:1997,Brockway:1996,Grifols:1996,Raffelt:1999,Rosenberg:2000,Horns:2012,Zioutas:2005,
Lamoreaux:2006,Yao:2006,Jaeckel:2007,Andriamonje:2007,Robilliard:2007,Zavattini:2008,Rubbia:2008}.

In this work we employ the latest distance duality relation (DDR) limits observed by Holand \& Barros \citep{Holanda:2016} 
to constrain the axion-photon coupling \cite{Bassett:2004,Bassett:2004b,Kunz:2004,Avgoustidis:2010}. 
We perform full $3D$ simulations for ALP-photon mixing in expanding $\Lambda$CDM 
universe, particularly for the galaxy clusters used in  Holand \& Barros \citep{Holanda:2016} analysis. We 
compare the simulation results with observational DDR limits and constrain coupling constant $g_{\phi}$. 
The coupling depends on background magnetic field and so are the results. Therefore, we effectively constrain the background magnetic field times 
the coupling. 

The axion-photon coupling framework is almost the same as in references \citep{Agarwal:2012,Tiwari:2012ax}. 
The background magnetic field is simulated in $3D$ space on a $512^{3}$ cubic grid, covering the comoving space 
up to $3.5$Gpc ($z=1$, all sources are within this distance). The origin of background magnetic field is assumed 
to be primordial \cite{Subramanian:2003sh,Seshadri:2005aa,Seshadri:2009sy,Jedamzik:1998,Subramanian:1998} and it 
has a power law spectrum in k-space. The background axion density is negligible (zero in simulation) 
as compared to photon flux emitted by source. 

The paper is organized as follows. In Section \ref{sc:Method} we give the 
estimate of axion-photon mixing on DDR relation. We briefly review the axion-photon mixing in Section 
\ref{sc:mixing}. In Section \ref{sc:sim} we provide the magnetic field generation and mixing simulation details. 
We present the observed limits in Section \ref{sc:result}. Finally, in Section \ref{sc:dis} we conclude and 
compare our results with other present limits. 
\section{Method}
\label{sc:Method}
The reciprocity theorem for null geodesics states that the geometric properties are 
invariant when we  exchange source and observer \citep{Etherington:1933,Ellis:2007}. 
Using this fundamental argument,  Etherington  in 1993 \citep{Etherington:1933} pointed out 
the distance duality relation (DDR), a relation between the luminosity distance $\dl$ and the 
angular diameter distance $\da$. The equation relating these two distances follows: 
\begin{equation}
\label{eq:ddr}
\dl = (1+z)^{2} \da
\end{equation}

The above equation (\ref{eq:ddr}) is quite general and true in any general Riemannian spacetime,  
however, requires that the source and observer are connected by null geodesics and the number of 
photons remains conversed. In principle the DDR relation can be tested observationally, if one can locate 
a source with well defined size and intrinsic luminosity. Interestingly, now the observational astronomy 
is becoming so efficient that by using different astronomical quantities several authors have attempted 
to test DDR relation. The validity of DDR relation assumes the photons conservation, however, in 
the presence of intergalactic magnetic field the coupling of photons with axion may potentially convert 
photons to axions and vice versa --this violates the photon number conversation. The mixing introduces 
the spread as well as deviation in DDR relation. The observed spread can be used to constrain the axion-photon 
mixing. 

The redshift dependent deviation of DDR relation is expressed as, 
\beq
\label{eq:eta1}
\frac{\dl}{(1+z)^2 \da} =\eta(z),
\eeq
we have $\eta(z)=1$ for strict DDR relation. Holanda et al. \citep{Holanda:2010} parametrized 
the redshift dependence of $\eta(z)$ in two distinct forms, $\eta(z)= 1+ \ezero z$(P1)  
and $\eta(z)= 1+ \ezero z/(1+z)$(P2) and investigated the $\ezero$ parameter 
by employing the luminosity distance $\dl$ measurements from Type Ia supernovae (SNe Ia) and diameter distance 
$\da$ from galaxy clusters \citep{Filippis:2005,Bonamente:2006}. Several other authors have also tested 
the DDR relation using different observations: SNe Ia plus cosmic microwave background  (CMB) and 
barion acoustic oscillations (BAO) \citep{Lazkoz:2008}, SNe Ia plus $H(z)$ data 
\citep{Avgoustidis:2009, Avgoustidis:2010,Holanda:2013,Liao:2015}, gas mass fraction of galaxy clusters and 
SNe Ia \citep{Holanda:2012, Goncalves:2012}, CMB spectrum \citep{Ellis:2013}, gamma-ray burst (GRB) plus 
$H(z)$ \citep{Holanda:2014}, SNe Ia plus BAO \citep{Puxun:2015}, gas mass fraction plus $H(z)$ \citep{Santos:2015}, 
gravitational lensing plus SNe Ia \citep{Holanda:2016SNeI}, SNe Ia and  radio galaxy plus CMB \citep{Rana:2016}.
Most of the above authors obtain no significant deviation in DDR relation, although, roughly the scatter in 
$\ezero$ parameter is observed as $\pm0.1$ to $\pm0.3$. Recently, Holand \& Barros \citep{Holanda:2016} test the 
DDR relation with $\da$ measurements from galaxy clusters \citep{Filippis:2005} plus $\dl$ measurements from 
SNe Ia. They report $\ezero=0.069\pm0.106$ (with parametric form P1) and $\ezero=0.097\pm0.152$ (P2) with their 
method I and  $\ezero=-0.0\pm0.135$ (P1) and $\ezero=-0.03\pm0.20$ (P2)  with their method II.  Several 
other measurements of $\ezero$ for the form P1, and P2 are also available in literature viz. Li et al. 
\citep{Li:2011} found $\ezero=-0.07\pm0.19$ (P1) and  $\ezero=-0.11\pm0.26$ (P2), Meng et al. \citep{Meng:2012}report 
$\ezero=-0.047\pm0.178$ (P1) and  $\ezero=-0.083\pm0.246$ (P2).

In axion-photon coupling scenario the photons can convert to axions (and vice-versa) and the number of 
photons remains no more conserved. The observed flux and the luminosity relation is expressed as:
\beq
\label{eq:flux1}
F = \frac{L}{4\pi \dl^2} \Rightarrow \dl \propto \frac{1}{\sqrt{F}}.
\eeq 

Assuming that the DDR is exact and the scatter is all due to axion-photon mixing, we can write $\eta(z)$
as, 
\beq
\label{eq:eta2}
\eta(z)= \frac{\dlobs}{\dl}= \sqrt{\frac{F}{F^{\rm obs}}},
\eeq
here $\dlobs$ and $F^{\rm obs}$ are the observed luminosity distance and flux respectively.  
Equation (\ref{eq:eta2}) for $\eta(z) \rightarrow 1$  can be approximated as, 
\beq
\label{eq:eta0}
\eta(z)-1= \frac{\dlobs -\dl}{\dl} \approx \frac{1}{2}\frac{F- F^{\rm obs}}{{F^{\rm obs}}}.
\eeq

If the redshift $z$ is known precisely then the scatter in $\ezero$ (times $z$ for P1 and $z/(1+z)$ 
for P2) is simply the scatter in $\frac{1}{2}\frac{F- F^{\rm obs}}{{F^{\rm obs}}}$. 
Alternatively,  the observed $\ezero$ scatter constrains the flux scatter due to photon-axion 
mixing or in other words the coupling of axions with photons. 
\section{Axion-photon mixing}
\label{sc:mixing}
\subsection{Mixing model}
The coupling of electromagnetic radiation with axions or ALPs  in presence of external magnetic field in 
flat expanding universe can be written as \cite{Carroll:1991,Garretson:1992,Tiwari:2012ax},
\begin{IEEEeqnarray}{rCl}
S= \int d^{4}x \sqrt{-g}~ \Big[-\frac{1}{4}F_{\mu\nu}F^{\mu\nu}-\frac{1}{4} g_{\phi} \phi F_{\mu\nu}\tilde{F}^{\mu\nu}
\nonumber\\+\frac{1}{2}(\omega_{p}^{2}a^{-3})A_{\mu}A^{\mu}+\frac{1}{2}g^{\mu\nu}\phi_{,\mu}\phi_{,\nu}-
\frac{1}{2}m^{2}_{\phi}\phi^{2}\Big],
\label{eq:S_expanding}
\end{IEEEeqnarray}
where $F_{\mu\nu}$ and $\tilde{F}^{\mu\nu}$ are electromagnetic field tensor  and dual tensor respectively, 
$\phi$ represents the pseudoscalar axion field, $g_{\phi}$  the  coupling between $\phi$  to electromagnetic field, 
$\omega_{p}= \frac{4\pi \alpha n_e}{m_e} $ is the plasma frequency (while $n_e$ and $m_e$ are the electron number 
density and mass respectively), `$a$' the cosmological scale factor, 
$A^{\mu}$ the electromagnetic four-potential and $m_\phi$ is the axion mass.  
The Maxwell's equations following this action are given in \citep{Agarwal:2012}. 
Assuming the z-axis as the photon propagation direction, the mixing equation in expending 
Universe is written as,   
\beq
\label{eq:mixing}
(\omega^{2}+ \partial^{2}_{z})\left(\begin{array}{c}
{\cal A}_{\parallel}\\ \chi \end{array} \right) -
M \left( \begin{array}{c} {\cal A}_{\parallel}\\ \chi \end{array} \right) =0, 
\eeq
where  $\omega$ is the radiation frequency and we have replaced $\phi$ by $\frac{\chi}{a}$.
${\cal A}_{\parallel}$ ($ {\vec{\cal A}}= \frac{(a^{2}{\vec E})}{\omega}$ ) refers to the 
component parallel to transverse magnetic field ($\mathcal{B}_\perp$) and  the mixing matrix,  
$M$, in above equation (\ref{eq:mixing}) has the following form, 
\beq
\label{eq:matrix} 
M = \left(\begin{array}{cc}

         \frac{\omega^{2}_{p}}{a}     & ~~~ -\frac{g_{\phi}}{a^{2}}
(a^{2}\mathcal{B}_\perp)\omega\\
-\frac{g_{\phi}}{a^{2}}\ (a^{2}\mathcal{B}_\perp)\omega   & ~~~ m^{2}_{\phi}a^{2} \end{array}\right).
\eeq

The solution to mixed field equation (\ref{eq:mixing}) is described in Ref. 
\cite{Das:2004ee,Agarwal:2008ac,Agarwal:2012},  we briefly review the same. 
The mixing matrix $M$ is diagonalized by an orthogonal transformation $OMO^{T}=M_{D}$, where 
 \beq O = \left(\begin{array}{cc}
         \cos\theta  &~~~  -\sin\theta\\
         \sin\theta  & ~~~  \cos\theta \end{array} \right), 
\eeq 
and $\theta$ is such as  $\tan2\theta ={2g_{\phi}\omega a^{-2} (a^{2}\mathcal{B}_\perp)}/{\left(\frac{\omega^{2}_{p}}{a}-m^{2}_{\phi}a^{2}\right)}$. The eigenvalues, $\mu_\pm$, of matrix $M$ are given as, 
\beq
\mu^2_\pm =\frac{\frac{\omega^{2}_{p}}{a} + m^{2}_{\phi}a^{2}} {2}\pm \frac{1}{2}\sqrt{\left(\frac{\omega^{2}_{p}}{a} + m^{2}_{\phi}a^{2}\right)^2 + (2g_{\phi}\omega \mathcal{B}_\perp)^2}
\eeq
\subsection{Modification to photon flux}
The mixing as described in above section modifies the electromagnetic radiation. The propagation of photons 
in mixing scenario in presence of external magnetic field is described in Ref. \citep{Agarwal:2008ac,Tiwari:2012ax}. 
Let the initial photon and pseudoscalar field $\chi$ ($a\phi$) density be represented as,
\beq
\rho(0) = \left( \begin{matrix} 
\langle{\cal A_{\parallel}}(0){\cal A_{\parallel}^{\ast}}(0)\rangle    & \langle{\cal A_{\parallel}}(0){\cal A_{\perp}^{\ast}}(0)\rangle   & \langle{\cal A_{\parallel}}(0) \chi^{\ast}(0)\rangle \\
\langle{\cal A_{\perp}}(0){\cal A_{\parallel}^{\ast}}(0)\rangle    & \langle{\cal A_{\perp}}(0){\cal A_{\perp}^{\ast}}(0)\rangle  & \langle{\cal A_{\perp}}(0) \chi^{\ast}(0)\rangle \\
\langle\chi(0) {\cal A_{\parallel}^{\ast}}(0)\rangle    & \langle\chi(0){\cal A_{\perp}^{\ast}}(0)\rangle   & \langle\chi(0) \chi^{\ast}(0)\rangle
\end{matrix} \right) 
\eeq

The modification in initial density matrix $\rho(0)$ after propagating distance $z$ can be written as, 
\beq
\label{eq:rhoz}
\rho(z)= P(z)\rho(0)P(z)^{-1},
\eeq
where $P(z)$ is an unitary matrix containing the detailed mixing solution. The full expression of $P(z)$ is
as following, 
\begin{widetext}
\beq 
P(z) = e^{i(\omega+\triangle_{A})z}\left(\begin{array}{ccc}
1-\gamma sin^2 \theta            &   0                  &         \gamma cos\theta sin\theta \\
0                                &  e^{-i[\omega+\triangle_A -(\omega^2- \omega^2_p)^{1/2}]z}  &          0 \\
\gamma cos\theta sin\theta        &     0                       &      1- \gamma cos^2\theta \end{array}\right).
\eeq
\end{widetext}
where $\gamma=1-e^{i\triangle z}$ , while  $\triangle =\triangle_\phi -\triangle_A$ and $\triangle_\phi$ , $\triangle_A$ 
are defined as ,
\beq \triangle_A = \sqrt {\omega^2-\mu^2_{+}}-\omega, ~~~~~~
\triangle_\phi = \sqrt{\omega^2-\mu^2_{-}}-\omega. \eeq
The first two row diagonal term of density matrix $\rho(z)$ represent the photon intensity 
after propagating distance $z$, 
\beq
\label{eq:flux2}
I(z)= \langle{\cal A_{\parallel}}(z){\cal A_{\parallel}^{\ast}}(z)\rangle + \langle{\cal A_{\perp}}(z){\cal A_{\perp}^{\ast}}(z)\rangle.
\eeq
Other terms in $\rho(z)$  represent the photon polarization and pseudoscalar field intensity. The probability producing an ALP from 
a photon after traveling through a distance $L$ while neglecting the fluctuations of background magnetic field and plasma density, 
is approximately given as \citep{Carlson:1994,Jain:2002vx}
\beq
P_{\gamma \rightarrow \phi} \approx  (g_{\phi} a \mathcal{B}_\perp l )^2 \sin^2 (L/2l), 
\eeq
where $l={2\omega a^{-1} }/{\left(\frac{\omega^{2}_{p}}{a}-m^{2}_{\phi}a^{2}\right)}$, and $L\gtrsim l$. For the visible $\omega$ and 
negligible ALP mass $\left(\frac{\omega^{2}_{p}}{a} \gg m^{2}_{\phi}a^{2}\right)$, $l$ is a few Mpc. In the case where 
$\frac{\omega^{2}_{p}}{a}$ close to $m^{2}_{\phi}a^{2}$ the conversion probability could be large. 

\section{Simulation details}
\label{sc:sim}
\begin{figure}[!t]
    \includegraphics[width=3.5in,angle=0]{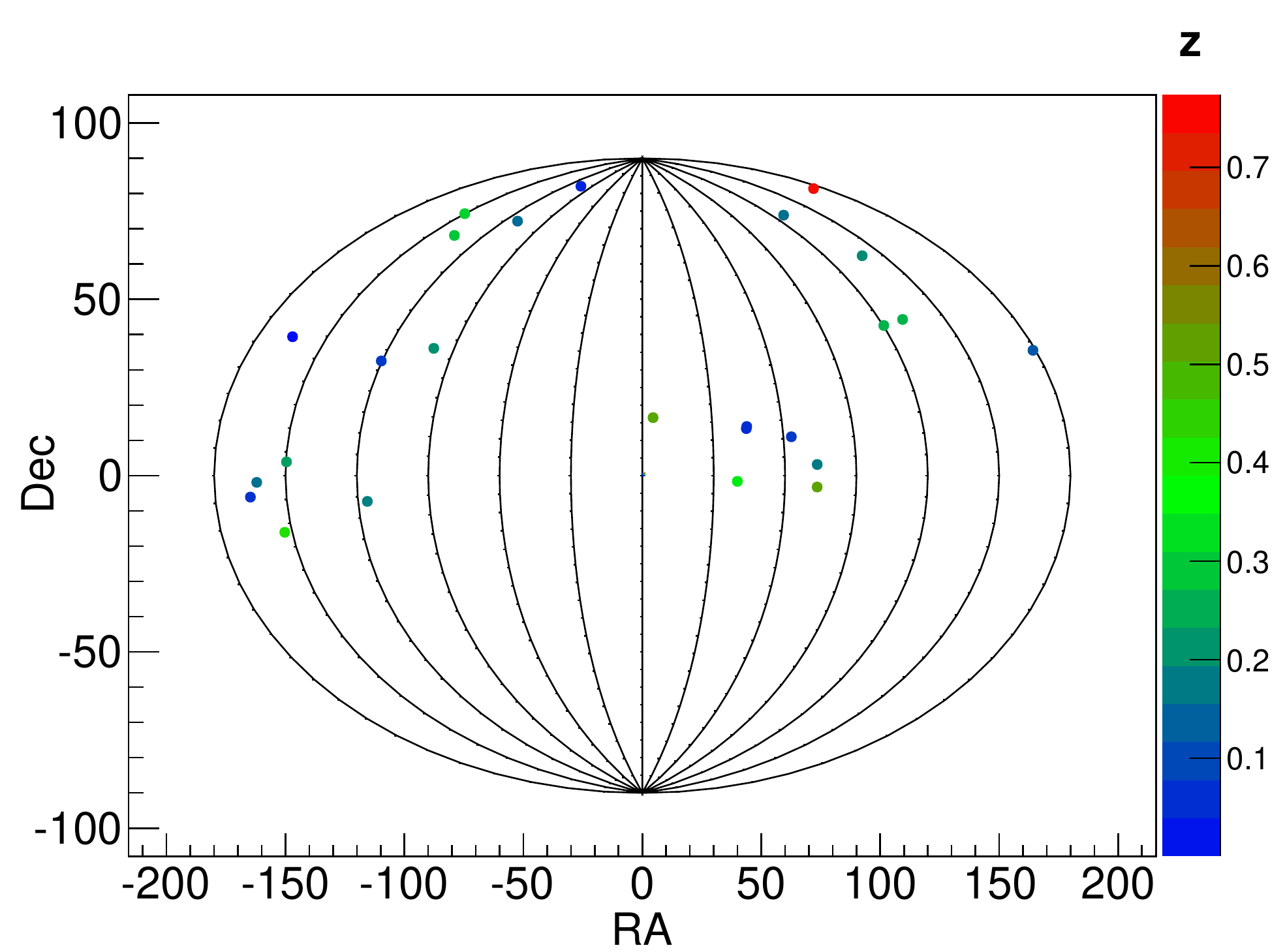}
 \caption{Redshifts and positions of De Filippis et al. \cite{Filippis:2005} galaxy clusters. 
We preserve the redshift distribution of these sources in our simulation.}
\label{fig:red}
\end{figure}

We simulate the axion-photon coupling for sources used in DDR validity tests by Holanda \& Barros \citep{Holanda:2016} 
in presence of external magnetic field.  De Filippis et al. \cite{Filippis:2005} provide the measurements of angular diameter 
distance ($\da$)  for 25 clusters as a function of redshift (see figure \ref{fig:red});
Holanda \& Barros \citep{Holanda:2016} employ the same for DDR validity test. We fix the initial source position as according 
to De Filippis et al. sample. To simulate axion-photon mixing we divide $3D$ space in $512^{3}$ cubic grids, covering 
the comoving space up to redshift 1, all the sources in De Filippis et al. sample are within this distance. The grid size in our 
simulation is $\sim13$ Mpc. 

To generate background magnetic field we assume it to be 
primordial\cite{Mack:2002,Subramanian:2003sh,Seshadri:2005aa,Seshadri:2009sy,Jedamzik:1998,Subramanian:1998}. 
The k-space correlation of homogeneous and isotropic magnetic field is expressed as,
\bea
        \langle b_{i}({\boldsymbol k}) b^{*}_{j} ({\boldsymbol q}) \rangle & = & \delta_{{\boldsymbol{k,q}}} P_{ij}({\boldsymbol k}) M(k) 
\label{eq:corr}
\eea
where $b_{i,j}({\boldsymbol k})$ {\footnote{%
The real space and k-space field transformations are as following:
\bea
\label{eq:IFT}
B_{j}({\boldsymbol r}) &=& \frac{1}{V(2\pi)^{3}} \sum b_{j}({\boldsymbol k}) e^{i {\boldsymbol k}.{\boldsymbol r}},\\
b_{j}({\boldsymbol k}) &=&\frac{1}{V} \sum B_{j}({\boldsymbol r}) e^{-i {\boldsymbol k}.{\boldsymbol r}}.
\eea
}}
are the  $i,j^{\rm th}$ component of the
magnetic field in k-space, $P_{ij}({\boldsymbol k}) = \left(\delta_{ij} -\frac{k_{i} k_{j}}{k^2}\right)$ 
is the projection operator and $M(k)=A k^{n_{_B}}$ contains the power-law nature of spectrum while  
$n_{_B}$  is power spectral index and  constant $A$ is normalization. We fix spectral index $n_{_B}=-2.37$ in our simulation. 
The results slightly depends on spectral index, however, the final effect on photon flux and polarization is expected to be 
small \citep{Tiwari:2012ax}. The normalization factor $A$ is set to fix real space 
magnetic field $B_{i}({\boldsymbol r})$, such as $\sum_i <B_i({\boldsymbol r}) B_i({\boldsymbol r})> = B_0^2$ , 
where $B_0$ is the strength of the magnetic field in real space averaged over a distance $r$,  
we set  $B_0=1$ nG \cite{Yamazaki:2010nf,Seshadri:2009sy} 
for a comoving scale of 1 Mpc. There is no k-space cutoff on correlation in our formulation and so in real space the 
correlation have no distance cutoffs. We first generate the k-space magnetic field  in each grid according to 
spectral distribution as in equation (\ref{eq:corr}). In k-space the grids are uncorrelated, for any given ${\boldsymbol k}$, 
$b_k=0$ and $b_\theta$, $b_\phi$ are generate independently from a Gaussian distribution 
\cite{Agarwal:2012,Agarwal:2011,Tiwari:2012ax} as, 
\begin{eqnarray}
\label{eq:gauss}
        f(b_{\theta}({\boldsymbol k}),b_{\phi}({\boldsymbol k})) = N \ {\rm exp} 
\left[-\left(\frac{b_{\theta}^{2}({\boldsymbol k}) + b_{\phi}^{2}({\boldsymbol k})}{2M({\boldsymbol k})}\right) \right],
\end{eqnarray}
where $N$ is a normalization factor. We Fourier transform the k-space field and obtain the real space magnetic field in
each grid. 

\begin{figure}[!t]
    \includegraphics[width=3.5in,angle=0]{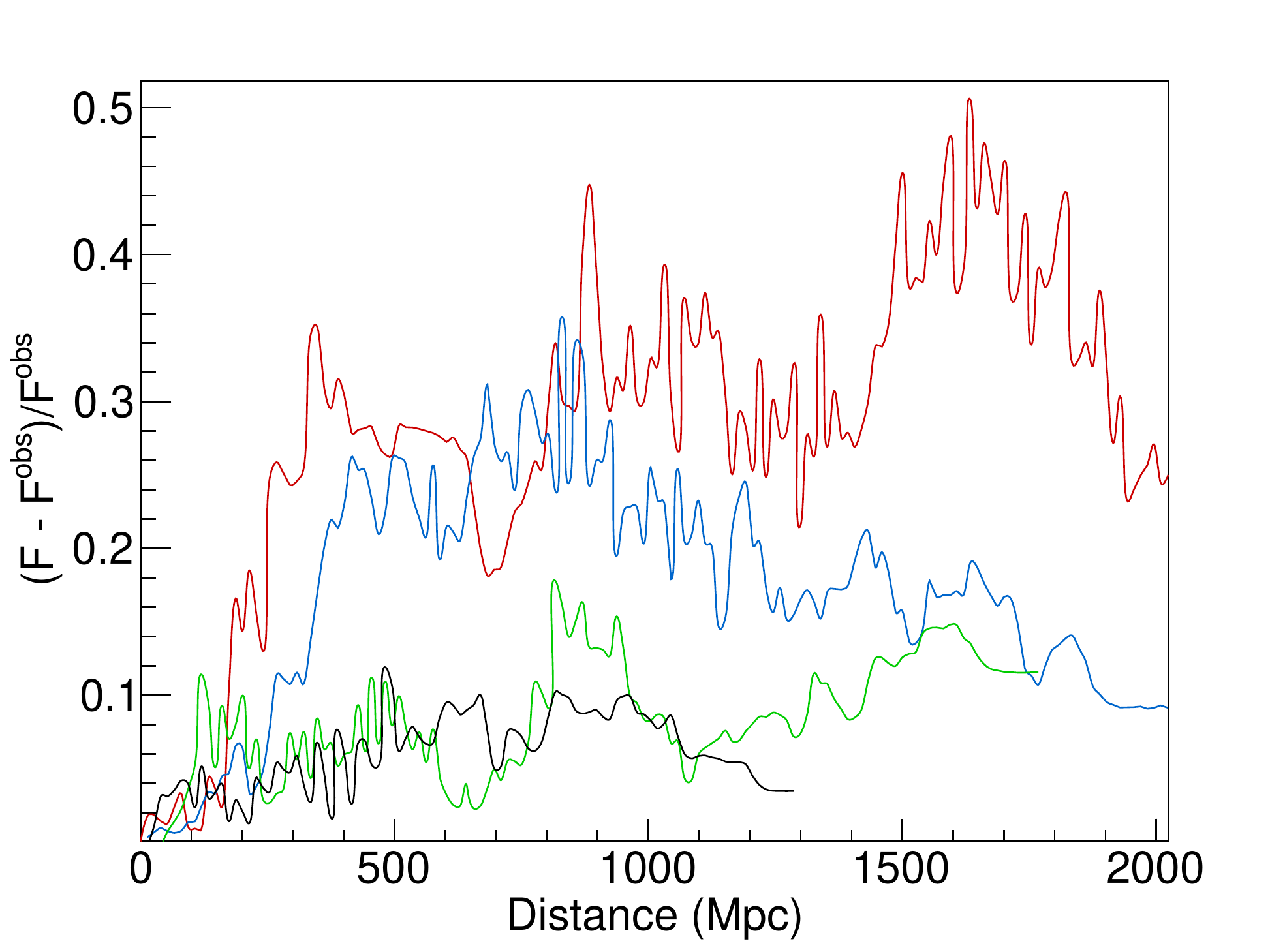}
    \caption{ The variation of photon flux along the line of sight for a few random realizations from our simulation. 
The origin in this plot represents the source location with initial flux $F$ equal to $F^{\rm obs}$, while propagation the 
photon flux fluctuate due to photon-ALP conversion. The lines are terminated once the light reaches to us. 
The termination point in figure represents the distance of the source from us and the final observed flux deviation.}
\label{fig:line}
\end{figure}

Having generated the background magnetic field in each grid, we fix the sources as in De Filippis et al. sample and propagate 
to us, the observer at central grid. The initial axion density is set to zero, the plasma density  $n_{e}$ is 
fixed to $10^{-8} a^{-3}{\rm cm}^{-3}$ and the radiation wavelength is set to visible (2 eV).  We have a fix 
coordinate (comoving distances)  system in our simulation. In order to propagate through each grid, we first rotate the coordinate 
so that the transverse component of the magnetic field aligns along one of the coordinate axes. We then use equation (\ref{eq:rhoz})
to calculate the modification due to mixing after propagating the particular grid. We then rotate back to resume our fixed coordinate 
system.  This procedure is repeated for all grids along the line of sight of a source. We have presented a few realizations from our 
simulation in figure \ref{fig:line}. The fluctuation of photon flux due to ALP-photon mixing is evident in figure.

We average over magnetic field seed by generating 400 random angular positions for each individual source while keeping its radial 
distance (redshift) fix. Each individual line of sight can be thought as a magnetic field configuration. Averaging over these 
400 random angular positions (line of sights) yields an average over magnetic field configurations. 
In total we simulate 10,000 (25 clusters$\times400$) intensity 
deviations ($\frac{1}{2}\frac{F- F^{\rm obs}}{{F^{\rm obs}}}$).   
\section{Results}
\label{sc:result}
\begin{figure}[!t]
    \includegraphics[width=3.5in,angle=0]{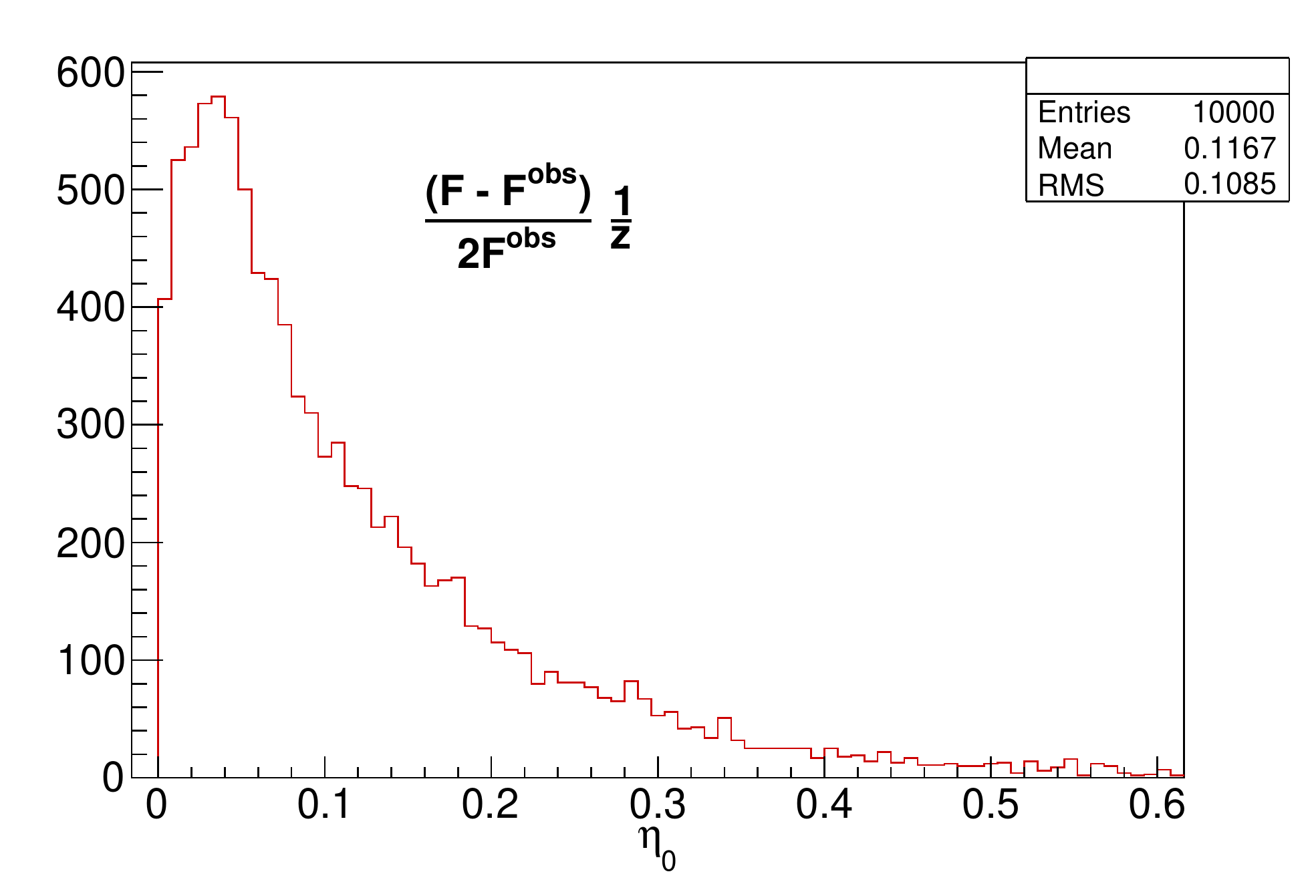}
    \caption{$\eta_0$ scatter following P1 distribution. The axion mass is set to $10^{-15} \gev$ and 
axion-photon coupling $g_{\phi}$ is set to  $6\times10^{-13} \gev^{-1}$. The RMS scatter is 0.1167 which is 
almost the Holanda \& Barros \citep{Holanda:2016} observed $\eta_0$ scatter. We have averaged over magnetic field 
seed by randomly generating the 400 angular positions for each individual source (25 clusters $\times 400$).}
\label{fig:P1dist}
\end{figure}

\begin{figure}
    \includegraphics[width=3.5in,angle=0]{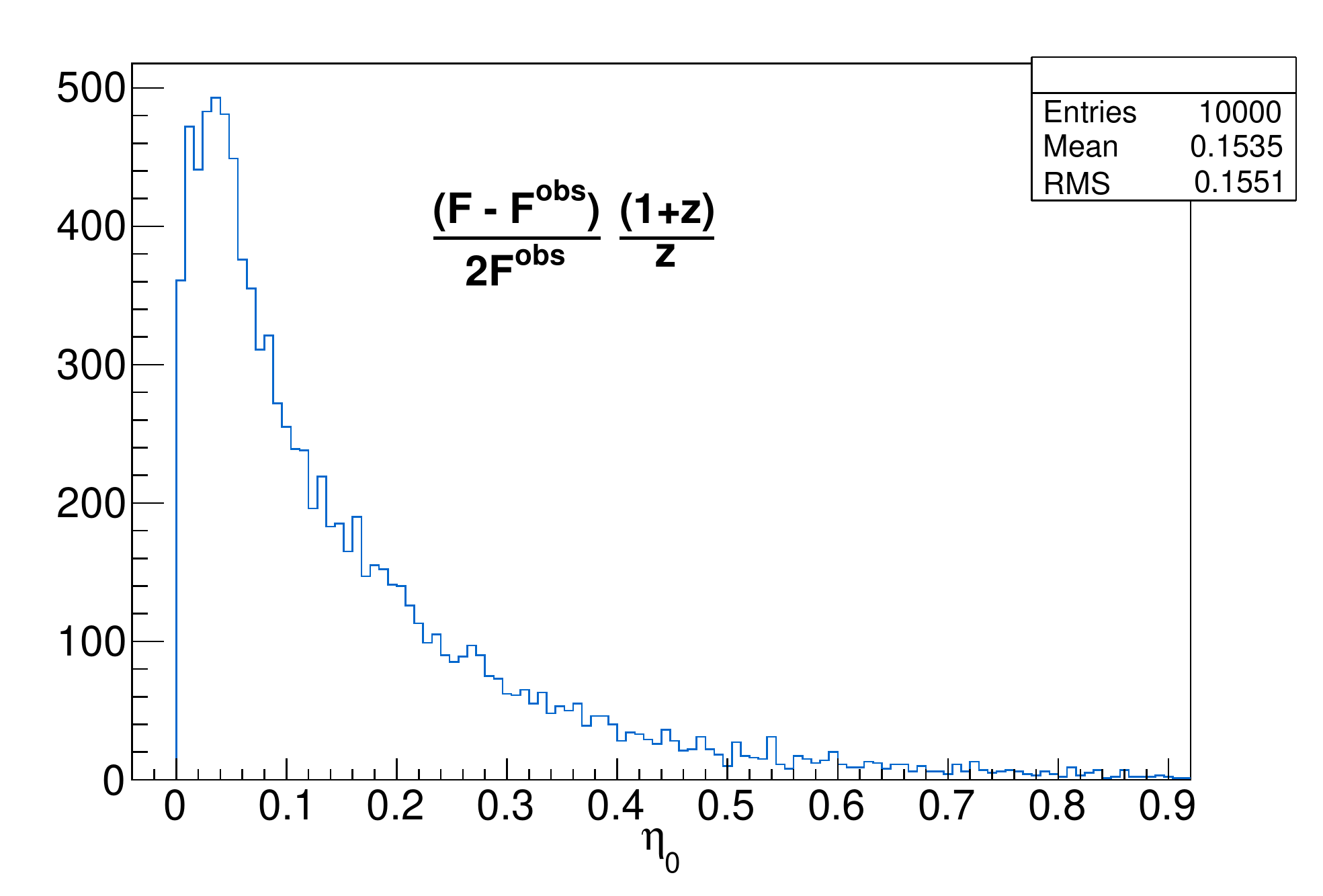}
    \caption{$\eta_0$ scatter following P2 distribution. The axion mass and axion-photon coupling 
values are same as in figure \ref{fig:P1dist}. Note that the RMS scatter is 0.1535 which is again almost the 
Holanda \& Barros \citep{Holanda:2016} observed $\eta_0$ scatter for P2.}
\label{fig:P2dist}
\end{figure}

We obtain ($\eta(z) -1$) from the fractional change in initial density to observed intensity (see equation 
(\ref{eq:eta0})). The critical scatter observed due to axion-photon mixing  for polynomial form P1 and P2 is shown in 
figure \ref{fig:P1dist} and \ref{fig:P2dist} respectively. We have tuned the coupling constant 
$g_{\phi}=6\times10^{-13}\gev^{-1}$ with axion mass $10^{-15} \gev$ to obtain root mean square (RMS) 
scatter in  ($\eta(z) -1$) within DDR observed limits as obtained by 
Holanda \& Barros \citep{Holanda:2016}. This is done by performing simulations for discrete values of $g_{\phi}$ in intervals of 
$10^{-13}\gev^{-1}$, this interval is small, we expect larger uncertainties in  $g_{\phi}$ due to assumed background 
magnetic field and plasma density. The mean value of $\ezero$ is not crucial 
as various DDR tests in literature give $\ezero$ ($\propto \eta(z) -1$) consistent with zero, 
namely Li, Wu \& Yu \citep{Li:2011} find $\ezero=-0.07\pm0.19$ and 
$\ezero=-0.11\pm0.26$ respectively for P1 and P2; Meng et al. \citep{Meng:2012} found $\ezero=-0.047\pm0.178$ (P1) and 
$\ezero=-0.083\pm0.246$ (P2);  Holanda \& Barros \citep{Holanda:2016} found $\ezero=0.069\pm0.106$ (P1) and 
$\ezero=0.097\pm0.152$ (P2) with their method I and $\ezero=-0.00\pm0.135$ (P1) and $\ezero=-0.03\pm0.20$ (P2) with 
their method II. 

The mean value and the root mean square (RMS) scatter can change depending on initial axion density. Here we constrain 
our analysis to zero initial axion density, however, there are processes which can generate some axion 
flux \citep{Jain:2009,Manousos:2011} at the source and those axions can convert to photons during there propagation 
form source to observer. For the time being,  with available DDR validity limits the RMS scatter of $\ezero$ parameter 
constrains the axion-photon coupling $g_{\phi}\le6\times10^{-13} \gev^{-1}$ for axion mass less than $10^{-15} \gev$. This 
is assuming that there is no systematic and statistical error in Holanda \& Barros \citep{Holanda:2016} DDR validity limits, 
including the scatter from these procedural errors the  coupling $g_{\phi}$ limits will go even lower. 

\begin{figure}[!t]
    \includegraphics[width=3.5in,angle=0]{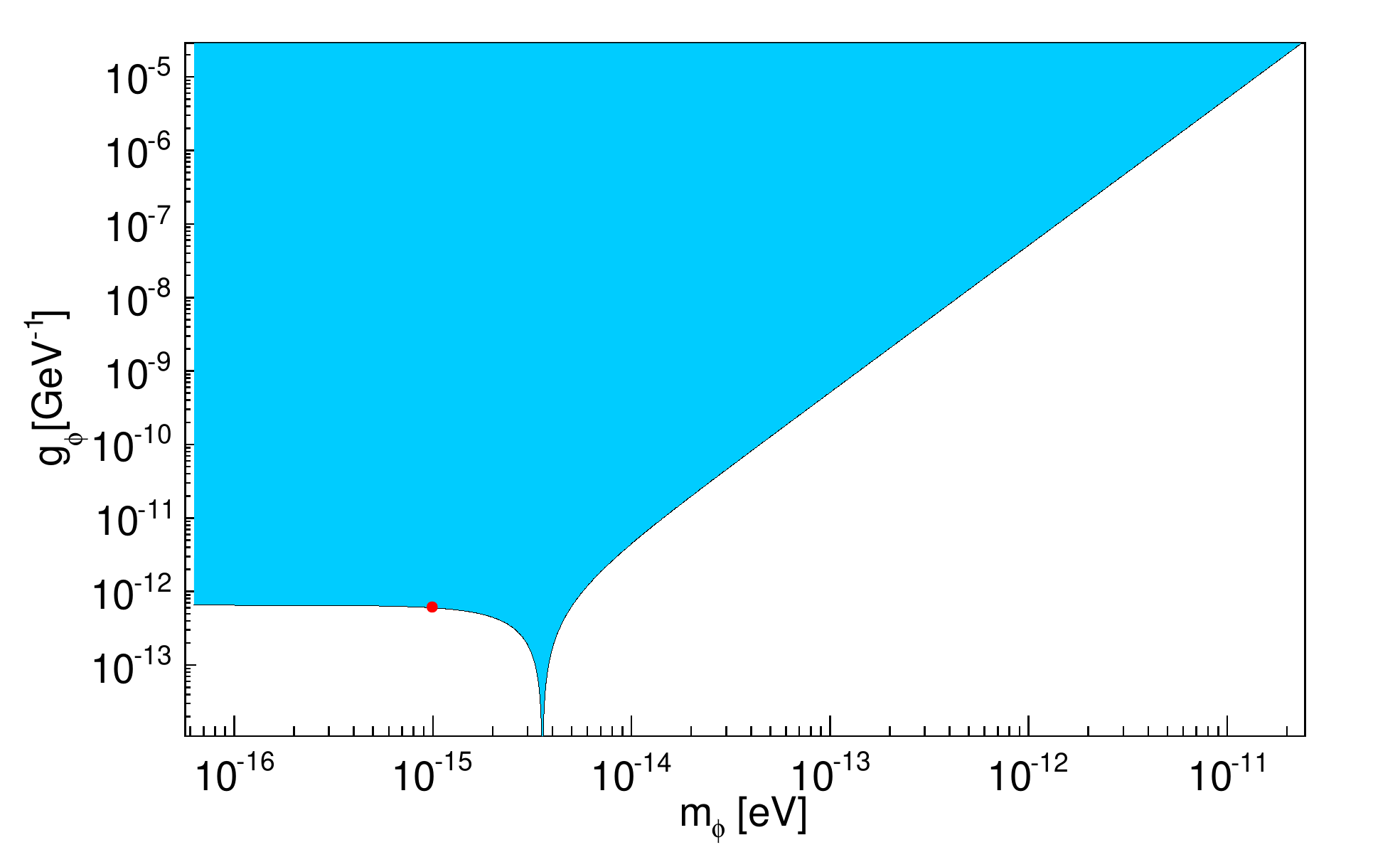}
    \caption{The extrapolated axion-photon mixing limits as a function of axion mass $m_{\phi}$ in static 
Universe ($a=1$). The strong dip between axion mass $10^{-15}$ - $10^{-14}$ eV is mixing resonance 
when axion mass becomes comparable to plasma frequency $\omega_p$. The simulation result in expanding Universe  obtained 
for mass $10^{-15} \gev$ is shown as red dot.  
}
\label{fig:limits}
\end{figure}

Finally, we extrapolate the $g_{\phi}$ limits  as a function of axion mass $m_\phi$ assuming the cosmological 
scale factor $a=1$ (static universe). The results are presented in figure \ref{fig:limits}, our simulation in expanding Universe is 
shown as a red dot in figure. The extrapolated value of $g_{\phi}$ for different axion mass $m_\phi$ may differ 
by a factor of 2 or 3 if compared with exact simulation in expanding universe. 

\section{Conclusion and Discussion}
\label{sc:dis}
In this work, we have presented the detailed simulation of ALP-photon mixing and constrained the axion-photon mixing 
from DDR validity observations. The basic principle of our method is the fact that photon conservation is violated in 
axion-photon mixing scenario, which is crucial to DDR validity \cite{Bassett:2004,Bassett:2004b,Kunz:2004,Avgoustidis:2010}. 
We employ the latest observation of DDR tests and present the limits from the observed scatter in relation. 
With present DDR validity observations, the mixing limit obtained are competitive with limit observed in Ref.\cite{Avgoustidis:2010}
and with other independent coupling observations. The results obtained in our work depends on background magnetic field strength and 
we constrain $g_{\phi}\le 6\times10^{-13} \gev^{-1}\left(\frac{nG}{\langle B \rangle_{\rm Mpc}}\right)$ 
for ALPs of mass $\le 10^{-15}$ eV. The ultra light ALPs are particularly interesting in astrophysics and can resolve 
major problems in galaxy formation \citep{Marsh:2014}.

The limits obtained in this work from DDR validity tests are much improved if compared with direct experiments i.e. 
using the CERN Axion Solar Telescope (CAST) Andriamonje et al. (2007) \citep{Andriamonje:2007} find 
$g_{\phi}< 8.8\times10^{-11} \gev^{-1}$ at 95\% CL for $m_\phi\le0.02$ eV.  The constraint on $g_{\phi}$ from Galactic 
Globular Clusters is $g_{\phi}<0.66\times10^{-10} \gev^{-1}$ \citep{Ayala:2014} at 95\% CL which is two orders of magnitude 
higher than this work. From X-ray observation the coupling $g_{\phi}\le 8.3\times10^{-12} \gev^{-1}$ for ALPs 
below mass  $7\times10^{-12}$ eV \citep{Wouters:2013}. 

The polarization measure of ultraviolet photons from active galactic nuclei yields a constraint 
$g_{\phi} B\le 10^{-11} \gev^{-1}$nG for ALPs with mass $\lesssim 10^{-15}$eV \citep{Horns:2012}. 
The polarization measurements from quasars give better constrain on $g_{\phi}$,  Payez et al. \cite{Payez:2012} found 
$g_{\phi}\le 2.5 \times 10^{-13} \gev^{-1}$ for ultra light ALPs. 

The CMBR polarization provide even more 
stringent constrain on coupling and  P. Tiwari \citep{Tiwari:2012ax} found
$g_{\phi}\le 1.6\times10^{-13}$ and  $3.4\times10^{-15}\gev^{-1} \left(\frac{nG}{\langle B \rangle_{\rm Mpc}}\right)$ 
for ALPs of mass $10^{-10}$ and $10^{-15}$ eV respectively. 

We conclude that the present DDR validity tests provide a good measure of axion-photon coupling limits. 
With time the DDR test will immensely improve and will limit the axion photon mixing drastically. 
Alternatively, the deviation observed in DDR validity test can be conventionally explained as 
axion-photon mixing, unless we have strong limits on coupling of ALPs with 
photon from some other observation or rule out the existence of ALPs somehow. 

\section*{Acknowledgements}
We thank Prof. Pankaj Jain and Rahul Kothari for a thorough reading of the manuscript.
This work is supported in part at the Technion by a fellowship from the Lady Davis Foundation.
We have used CERN ROOT 5.34/21 \citep{root} for generating our plots.
\bibliographystyle{prsty}
\bibliography{CMBR,cddr}
\end{document}

%% file: def.tex
\usepackage[ngerman,english]{babel}
\usepackage[latin9,utf8]{inputenc}
\newif\ifAMStwofonts
\AMStwofontstrue 

\def\gev{{{\rm GeV}}}
\def\ezero{{{\eta_{_0}}}}
\def\da{{{\rm D_A}}}
\def\dl{{{\rm D_{L}}}}
\def\dlobs{{{\rm D_{L}^{obs}}}}

\def\gsim{~\rlap{$>$}{\lower 1.0ex\hbox{$\sim$}}}

\def\simpropto{\lower.2ex\hbox{$\; \buildrel \propto \over \sim \;$}}
\def\ltsim{\lower.5ex\hbox{$\; \buildrel < \over \sim \;$}}
\def\gtsim{\lower.5ex\hbox{$\; \buildrel > \over \sim \;$}}
\def\ltsim{\lower.5ex\hbox{$\; \buildrel < \over \sim \;$}}
\def\gtsim{\lower.5ex\hbox{$\; \buildrel > \over \sim \;$}}

\def\pmb#1{\setbox0=\hbox{#1}%
\kern-.025em\copy0\kern-\wd0
\kern.05em\copy0\kern-\wd0
\kern-.025em\raise.0433em\box0}

\def\simlt{\lower.5ex\hbox{$\; \buildrel < \over \sim \;$}}
\def\simgt{\lower.5ex\hbox{$\; \buildrel > \over \sim \;$}}

\def\beqa{\begin{eqnarray}}
\def\eeqa{\end{eqnarray}}
\def\fixit#1{}

\newcommand{\beq}{\begin{equation}}
\newcommand{\eeq}{\end{equation}}
\newcommand{\ignore}[1]{}
\newcommand{\be}{\begin{equation}} \newcommand{\ee}{\end{equation}}
\newcommand{\bea}{\begin{eqnarray}} \newcommand{\eea}{\end{eqnarray}}